\documentclass[final,5p,times]{elsarticle}

\usepackage{graphicx}
\usepackage{epsf}
\usepackage{epsfig}   
\usepackage{cancel}

\usepackage{subfigure}
\usepackage{amsmath,amssymb}
\usepackage{rotating}
\usepackage{bm}
\usepackage{color}

\usepackage{verbatim}

\def\eq#1{(\ref{#1})}
\def\eqref#1{Eq.~(\ref{#1})}

\def\sec#1{Sec.~\ref{#1}}

\def\x{\vec{x}}

\newcommand{\be}{\begin{equation}}
\newcommand{\ee}{\end{equation}}

\newcommand{\avg}[1]{\left\langle#1\right\rangle}

\newcommand{\Tr}{\text{Tr}}

\def\g{g_0}

\newcommand{\sign}{\text{sign}}

\def\gth{f_0^2}
\def\gthsq{f_0^4}
\def\dlam{\tilde{\lambda}}
\def\lampar{\lambda_\parallel}

\begin{document}

\begin{frontmatter}

\title{Nonlinear sigma models at nonzero chemical potential:\\ breaking up instantons and the phase diagram}

\author{Falk Bruckmann}
\author{Tin Sulejmanpasic}
\address{Institute for Theoretical Physics, University of
  Regensburg, 93040 Regensburg, Germany}

\begin{abstract}
We consider asymptotically free nonlinear sigma models in two dimensions which, due to their internal symmetries, allow for a conserved charge. Introducing nonzero chemical potential for the \emph{SO(2)} subgroup of the symmetry group, we discuss two expected phase transitions, which are related to charge condensation and percolation of merons, respectively. The latter are topological objects with half integer charge similar to vortices in the abelianized $O(2)$ theory, that emerge for large chemical potentials due to the suppression of the complementary field components. We conjecture a particular ordering of these transitions supported by large $N$ calculations, and discuss dualities helpful for the understanding of these systems in the continuum and on the lattice. We suggest that a similar behaviour is to be expected in QCD at nonzero density.
\end{abstract}


\end{frontmatter}

\section{Introduction}

Since our main motivation are gauge theories, we discuss the relevant aspects of them first. Ever since the birth of Quantum Chromodynamics (QCD) and its high-energy perturbative successes, the low-energy physics, dominated by nonperturbative phenomena, remains elusive. Hence, two key phenomena characteristic for QCD, chiral symmetry breaking and confinement, are still not understood. 

Chiral symmetry breaking phenomena have been related to instantons due to their low-lying quark modes (for a review see \cite{Schafer:1996wv}), whereas confinement is usually attributed to monopole-like configurations\footnote{There have been attempts to use fractional instantons called \emph{merons}~\cite{Callan:1977gz}.}. In gauge theories without a Higgs field, monopoles have first been suggested to emerge through gauge fixing \cite{'tHooft:1981ht}. On space-times with compact directions like at non\-zero temperature, instantons reveal constituents \cite{Gross:1980br,Lee:1997vp,Kraan:1998pm,Lee:1998bb}, the so-called instanton-monopoles. There has been a lot of progress concerning their role in recent years, mainly in supersymmetric cousins of pure Yang-Mills (YM) theory \cite{Seiberg:1994rs,Davies:1999uw,Poppitz:2012sw,Poppitz:2012nz} or via including center-stabilizing terms \cite{Unsal:2008ch}. 

For equilibrium QCD at \textit{nonzero density}, which occurs in nuclear matter, neutron stars etc., the powerful nonperturbative methods of numerical lattice simulation are hampered by the infamous sign problem. Although properties at asymptotically high chemical potential \cite{Alford:1997zt,Alford:1998mk} and the influence of nonzero baryon densities on instanton-liquid like models were treated long ago \cite{Rapp:1999qa,Schafer:2000et}, an understanding of the interplay between confinement and chemical potentials of the order of the strong scale $\Lambda_{QCD}$ is still lacking.

Meanwhile, two-dimensional sigma models such as $O(N)$ and $CP(N-1)$ are very popular as toy models\footnote{The $O(2$) and $O(3)$ model are even more important for condensed matter systems.} of YM and QCD. They exhibit a variety of YM-like\footnote{Fermions can be coupled to sigma models as well. In our study we will not include them.}  behavior such as asymptotic freedom, mass gap generation and, in the case of ${CP(N-1)}$ models, instantons including fractional constituents \cite{Eto:2004rz,Eto:2006mz,Bruckmann:2007zh,Brendel:2009mp,Harland:2009mf}\footnote{which play a role in the cancellation of the ambiguities in the Borel resummation of the perturbative series \cite{Dunne:2012ae}} 
(fractionalized saddle points also occur in principal chiral models \cite{Cherman:2013yfa,Cherman:2014ofa} and $O(N)$ models \cite{Unsal_private}). Much simpler than their four-dimensional gauge theory counterparts, they offer an attractive playground for understanding nonperturbative dynamics. 

In this view it is interesting to note that the sigma models can be considered at nonzero density as well. Obviously, the $O(N)$ and $CP(N-1)$ models enjoy global symmetries, and it is possible to add chemical potentials $\mu$ for their conserved charges. Actually, it is easy to convince oneself that the boundary conditions of fractionalized $CP(N-1)$ instantons amount to imaginary chemical potentials.
With real chemical potential we are thus able to study an asymptotically free theory with a  dynamically generated mass gap at nonzero density: a theory that shares the so-called Silver-Blaze phenomenon and the sign problem with QCD, but with a potential benefit of large $N$ computability. 

Being a bosonic system, however, the chemical potential does not only couple linearly to the conserved charges, but also suppresses certain field components (quadratically, see below). In this work we focus on the $O(N$), most particularly the $O(3)$ model. Adding a large chemical potential for some $S\!O(2)\in O(N)$ breaks the symmetry down to this subgroup and makes the theory weakly coupled and prone to perturbative treatment, much like in QCD. A similar breaking can be achieved by an\-iso\-tro\-pies in the field space \cite{Hikami01021980,Ogilvie:1981yw,Satija:1982ja,Haldane:1982rj,Affleck:1985jy,Vachaspati:2011ad}. The chemical potential differs from the latter by the linear (purely imaginary) term, which removes the (Euclidean Lorentz) isotropy in time and space\footnote{There are also breakings using external magnetic fields \cite{Wiegmann:1985jt,Polyakov:1983tt,Schutzhold:2004xq}.}. 

Now the resulting $O(2)$ model, or the $XY$ model as its lattice counterpart, can undergo a (Berezinsky-)Kosterlitz-Thouless \cite{Berezinsky:1970fr,Kosterlitz:1973xp} (KT) transition. In the latter the two-point correlators change their decay from exponential to algebraic, although there is no order parameter (`infinite order' transition). It is the strongest order-disorder transition that can occur in two dimensions. This transition is known to be driven by nonperturbative objects: vortices, which in the case of $O(3)\rightarrow O(2)$ breaking carry topological charge $1/2$ and are dubbed `merons' \cite{Affleck:1988nt}. At large $\mu$ and hence small coupling the vortices come in tightly bound pairs. In this regime the correlators are algebraically decaying, whereas the vortices percolate and disorder the system at smaller $\mu$, with exponentially decaying correlators. This is reflected by the free energy dependence on the topological theta-angle, which should enter as $\cos\theta$ from vortex--anti-
vortex pairs 
(i.e.\ instantons) or as a function of $\cos(\theta/2)$, when vortices dominate. 

The vortices (or `merons') which appear in the effective $O(2)$ low energy theory, visit the poles at their cores in the full $O(3)$ theory, so that their action is non-singular even in the continuum limit. A pair of such merons, which has no winding at infinity and where positive topological charges add up to unity, can thus be considered as an instanton dissociated into two constituents. Due to these objects, there are potentially two phase transitions in the $O(3)$ model as a function of the chemical potential $\mu$: the KT transition at some $\mu_{KT}$ and the second order transition at the critical $\mu_c$, the end-point of the Silver-Blaze region, where the corresponding charge is induced in the system. The latter is intimately related to the energy of the lowest excitation, hence the dynamically generated mass. We will demonstrate this in a large $N$ calculation in $O(N)$. We conjecture that $\mu_c<\mu_{KT}$, since the KT transition cannot occur in the phase with full symmetry. Using a dualization to 
the Sine-Gordon model we will show how $\mu$ renders the two-point correlators in the temporal 
direction different from spatial ones. The KT transition then will take place in a space-time anisotropic regime and our arguments above shall apply to the spatial correlators. Unfortunately, the region in between $\mu_c<\mu<\mu_{KT}$ is not accessible to a semiclassical analysis. Nonetheless, the chemical 
potential has a nice interpretation in terms of 
the dual Sine-Gordon theory: it couples to the (topological) charge of kinks. We propose to 
study the details of this highly nontrivial system through lattice simulations and discuss lattice dualizations to low energy variables, that also eliminate the sign problem.

This letter is organized as follows. In the next section we define $O(N)$ and $CP(N-1)$ models, their conserved charges and chemical potentials. Thereafter, in Sec.~\ref{sec:O2} we discuss the $O(2)$ model including its dualization to Sine-Gordon and the anisotropy of correlators. In Sec.~\ref{sec:O3} we generalize to the $O(3)$ model and present our conjecture for its phase diagram. Sec.~\ref{sec:largeN} contains the large $N$ calculation and Sec.~\ref{sec:lattice} the lattice formulations. We conclude in Sec.~\ref{sec:summary}.

\section{Sigma models with chemical potential}
\label{sec:defs}

The (continuum) Lagrangian of the $O(N)$ model without chemical potential reads $\partial_\nu \bm n\, \partial^\nu \bm n/(2g_0^2)$
where $\bm n^2=1$, $\nu=(0,1)=(t,x)$ and $\g$ is the bare coupling. 
The system enjoys a global $O(N)$ symmetry, the $S\!O(N)$ subgroup\footnote{The remaining transformations are reflections and do not induce Noether charges.} of which gives rise to conserved currents and charges\footnote{Note that this is just the total $N$-dimensional angular momentum of vectors~$\bm n$.}
\be\label{eq:charge_def}
 Q^a=\frac{1}{\g^2}\int\! dx\, \bm n\cdot T^a\dot{\bm n}\;,
\ee
where $T^a$ are the antisymmetric traceless generators of $S\!O(N)$.
In the thermodynamic context the Euclidean Lagrangian is used and from now on $\dot{\bm n}$ stands for the derivative with respect to Euclidean time $x_0$. The introduction of a chemical potential coupling to these conserved charges is straightforward (e.g.\ via the Hamiltonian picture integrating out momenta), 
\begin{align}
 \g^2\mathcal L
 =&\,\frac{1}{2}\,\Big[(\partial_\nu-i\delta_{\nu 0}\mu^aT^a)\bm n\Big]^2
 \label{eq:Lfirst}\qquad(O(N))\\
 =&\,\frac{1}{2}(\partial_\nu\bm n)^2
 +i\mu^a{\bm n}\cdot T^a \dot{\bm n}
 +\frac{1}{2}\bm n\cdot (\mu^a T^a)^2\bm n\;.
 \label{eq:Lsecond}
\end{align}

We will focus on chemical potentials, whose generator $T^a$ acts on two components only (all other cases can be obtained by adding several such `fundamental' chemical potentials), which we can then choose to be $n_{1,2}$, in which $(T^a)_{bc}=(i\tau_2)_{bc}$ with $\tau_2$ the second Pauli matrix,
\begin{align}
 \g^2\mathcal L
 =\frac{1}{2}(\partial_\nu\bm n)^2+i\mu({n_1}\dot n_2-n_2\dot n_1)
 -\frac{1}{2}\,\mu^2 (n_1^2+n_2^2)\label{eq:Lagain1}\;.
\end{align}
The $n_{1,2}$-terms are those of a free scalar field $n=n_1+in_2$ subject to a $U(1)$ chemical potential (with charge proportional to $i(n^*\dot{n}-\dot{n}^* n)$). However, $n_{1,2}$ are normalized together with the remaining field components. The normalization can be used to rewrite, c.f.\ \cite{Hasenfratz:1990ab}
\begin{align}
 \g^2\mathcal L
 =&\,\frac{1}{2}(\partial_\nu\bm n)^2+i\mu({n_1}\dot n_2-n_2\dot n_1)
 +\frac{1}{2}\,\mu^2 {\bm n}_\perp^2-\frac{1}{2}\mu^2\label{eq:Lagain2}\;,
\end{align}
where we have split
\be
 {\bm n}^T=(\underbrace{n_1,n_2}_{\substack{{\bm n}_\parallel^T}},
 \underbrace{n_3,\ldots,n_N}_{\substack{{\bm n}_\perp^T}})\;.
 \label{eq:n_decomp}
\ee

In contrast to the fermionic case, in e.g.\ QCD, the chemical potential enters in two terms that cause different effects: The quadratic term obviously suppresses all but the two parallel components, explicitly breaking $O(N)$ to $O(2)$. A similar an\-iso\-tro\-py was introduced ad hoc into the $O(3)$ model by various authors \cite{Haldane:1982rj,Affleck:1985jy,Vachaspati:2011ad} and our system will also have `meron' configurations discussed by Affleck \cite{Affleck:1988nt} (see below). We have kept the constant term $-\mu^2/2$ because it will contribute (a constant) to the charge. The term linear in $\mu$ contains the charge and  distinguishes space and time. Since it is imaginary, the path integral weight is not positive anymore, which in numerical simulations causes a sign problem\footnote{unless purely imaginary chemical potentials are considered}.

The $CP(N-1)$ models are written in terms of an $N$-di\-men\-sio\-nal vector ${\bm u}$ normalized to $1$, ${\bm u}^\dagger {\bm u}=1$, with an action that can be written in a quadratic form by virtue of a composite hermitian gauge field that enters the antihermitian covariant derivative 
$D_\nu=\partial_\nu+i A_\nu$ or in a quartic form solely in ${\bm u}$. Again we focus on chemical potentials acting on the first two components, which is achieved by taking the corresponding hermitian generator of $S\!U(N)$ in the Cartan subalgebra, $\tau=\text{diag}(1,-1,0,\ldots,0)$,
\begin{align}
 \g^2\mathcal L
 =&\,\Big[(D_\nu+\delta_{\nu 0}\mu\tau){\bm u}\Big]^\dagger
 \Big[(D_\nu-\delta_{\nu 0}\mu\tau){\bm u}\Big]\qquad (CP(N-1))\\
 =&\,-{\bm u}^\dagger(D_\nu-\delta_{\nu 0}\mu\tau)^2\,{\bm u}\;,
 \label{eq:mu_in_cp}
\end{align}
where to $D_0$ a hermitian admixture proportional to $\mu$ is added, such that the action is not real. The quartic version is obtained by using the equation of motion for the gauge field 
$A_\nu
 =i{\bm u}^\dagger\big(\partial_\nu+\delta_{\nu 0}\mu\tau\big)\,{\bm u}$, and plugging it back into the covariant derivative in (\ref{eq:mu_in_cp}).

\section{$O(2)$: dualization and role of the chemical potential}
\label{sec:O2}

The $O(2)$ model can be written entirely in terms of an angle $\phi$ via ${n_1=\cos\phi}$, ${n_2=\sin\phi}$, the Lagrangian reads,
\be\label{eq:O2lag}
 \g^2\mathcal L
 =\frac{1}{2}(\partial_\nu\phi)^2+i\mu\,\dot{\phi}-\frac{1}{2}\,\mu^2\qquad (O(2))\;.
\ee
The $\beta$-function vanishes in perturbation theory, but nonpertubatively the compactness of the variable $\phi$ allows for \textit{vortices}. They are characterized by integer winding numbers $1/(2\pi)\cdot\oint_C dx_\nu\,\partial_\nu\phi$ on closed contours $C$ around their cores. To reveal the effects of these vortices, several dualization techniques have been developed for the $XY$-model, Eqs.~(\ref{eq:O2_lattice_naive},\ref{eq:O2_lattice_velocity}), as the lattice version of the $O(2)$ model \cite{Jose:1977gm}. For reasons of clarity we postpone lattice formulations to Sec.~\ref{sec:lattice} and first discuss the continuum version of the dualities, cmp.~e.g.\ \cite{Affleck:1988nt}, in close analogy to Polyakov's treatment of confinement in $2+1$-dimensional gauge theories \cite{Polyakov:1976fu}.

We introduce a velocity field $v_\nu=\partial_\nu\phi$, that is conservative in the absence of vortices, i.e.\ its curl (vorticity) vanishes, $\epsilon_{\nu\rho}\partial_\nu v_\rho=0$. We impose this constraint with a Lagrange multiplier field $\sigma(x)$ through a term
\be
 -\frac{i}{2\pi}\,\sigma\epsilon_{\nu\rho}\partial_\nu v_\rho=-\frac{i}{2\pi}\,v_\nu\epsilon_{\nu\rho}\partial_\rho \sigma
\label{eq:Lagrange_O2} 
\ee
(the normalization is a matter of convenience) in addition to the original Lagrangian \eqref{eq:O2lag} written in terms of $v$: $(v_\nu^2+2i\mu v_0-\mu^2)/(2g_0^2)$. This allows to integrate out the quadratically appearing $v_\nu$. We obtain an action for the Lagrange multiplier field, called dual field, only,
\be
 \mathcal L=\frac{g_0^2}{2(2\pi)^2}(\partial_\nu\sigma)^2-\mu\frac{\partial_x\sigma}{2\pi}\;.
\label{eq:action_O2_without}
\ee
Already this action has two remarkable properties. First, it is real even in the presence of chemical potential, whereas the original action for $\phi$ has an imaginary part (for the lattice analogue see \cite{Meisinger:2013zfa}). Secondly, the chemical potential couples to the zeroth component of the current $\epsilon_{\nu\rho}\partial_\rho\sigma$ that is conserved geometrically (without a Noether symmetry). That the corresponding charge is quantized will become clearer upon introducing vortices below. Likewise, the $\mu$-term does not enter the equations of motion for $\sigma$.

To compute the two-point function
\be 
 \avg{e^{-i\phi({\bm x}_1)}e^{i\phi({\bm x}_2)}}
 =\avg{e^{i\int_{{\bm x}_1}^{{\bm x}_2}\!dx_\nu v_\nu}}
\label{eq:two_point} 
\ee
in the dual field, one first has to couple $v_\nu$ to a (tangential) delta function on some contour $C$ connecting\footnote{Even when the conservativeness of $v$ is violated by vortices, this phase factor is independent of the choice of the contour $C$.} ${\bm x}_1$ and ${\bm x}_2$. Upon integrating out $v_\nu$ again, the new term in the action can be interpreted as a line of dipole sources on $C$ oriented perpendicular to it. In other words, the field $\sigma$ is required to have a jump of $2\pi$ across the contour $C$. Therefore, this field has to be compact with period
$2\pi$, too\footnote{Viewing $\sigma$ as a compact field may be a cause of some confusion, as one may think that this allows for ``dual vortices'', i.e.\ excitations for which the field wraps around some point $\bm x$. Due to this problem one can think of the $\sigma$ field as noncompact, and just require it to have appropriate discontinuous jumps by $2\pi$ when operator insertions $e^{\pm i\phi(x)}$ are considered. We however prefer to think of the field $\sigma$ as compact, but the space-time as having an appropriate UV definition which does not allow for the dual vortices. The compact nature of the sigma field is also important when thinking of the theory on a spatial tourus, where $\sigma$ wrappings in the spatial direction are interpreted as charged physical particles (see below).}.  

Now we come to the vortices. The integral $\oint_C dx_\nu\,v_\nu=\pm 2\pi$ for each contour $C$ around a unit vortex location ${\bm X}$ corresponds to a delta source at ${\bm X}$,
\be
 \epsilon_{\nu\rho}\partial_\nu v_\rho=\pm 2\pi\delta^{(2)}({\bm x}-{\bm X})\;.
\ee
This term has to be coupled to $\sigma$ in addition to \eqref{eq:Lagrange_O2}, which upon space-time integration becomes $\pm i\sigma({\bm X})$. In other words, an insertion of $e^{\pm i\sigma({\bm X})}$ in the path integral creates a (anti)vortex at ${\bm X}$. Then one has to integrate over all possible locations and charges of (anti)vortices with some fugacity, say $m^2$, which yields
\be
 \mathcal L=\frac{g_0^2}{8\pi^2}(\partial_\nu\sigma)^2-\mu\frac{\partial_x\sigma}{2\pi}
 -m^2\cos\sigma\;.
\label{eq:action_O2_with}
\ee
This is the celebrated two-dimensional Sine-Gordon model, now including a chemical potential. This system has (anti)kink solutions where $\sigma$ moves from one minimum of the potential at $x\to- \infty$ to its successor/predecessor at $x\to\infty$, in a transition region of size proportional to $1/(g_0 m)$. Then the integral $\int\! dx\,\partial_x\sigma/(2\pi)$ is obviously the kink number, that is conserved in time. It means that in the dual formulation the $\mu$ couples to a topological charge, like a theta-term, whereas in the original ${\bm n}$-theory it coupled to a Noether charge, hence particles have become kinks. A similar effect takes place in the equivalence of the Sine-Gordon to the Thirring model: in the latter the topological charge becomes a Noether charge of fermions.

\begin{figure*}[htbp] 
   \centering
   \includegraphics[width=0.75\textwidth]{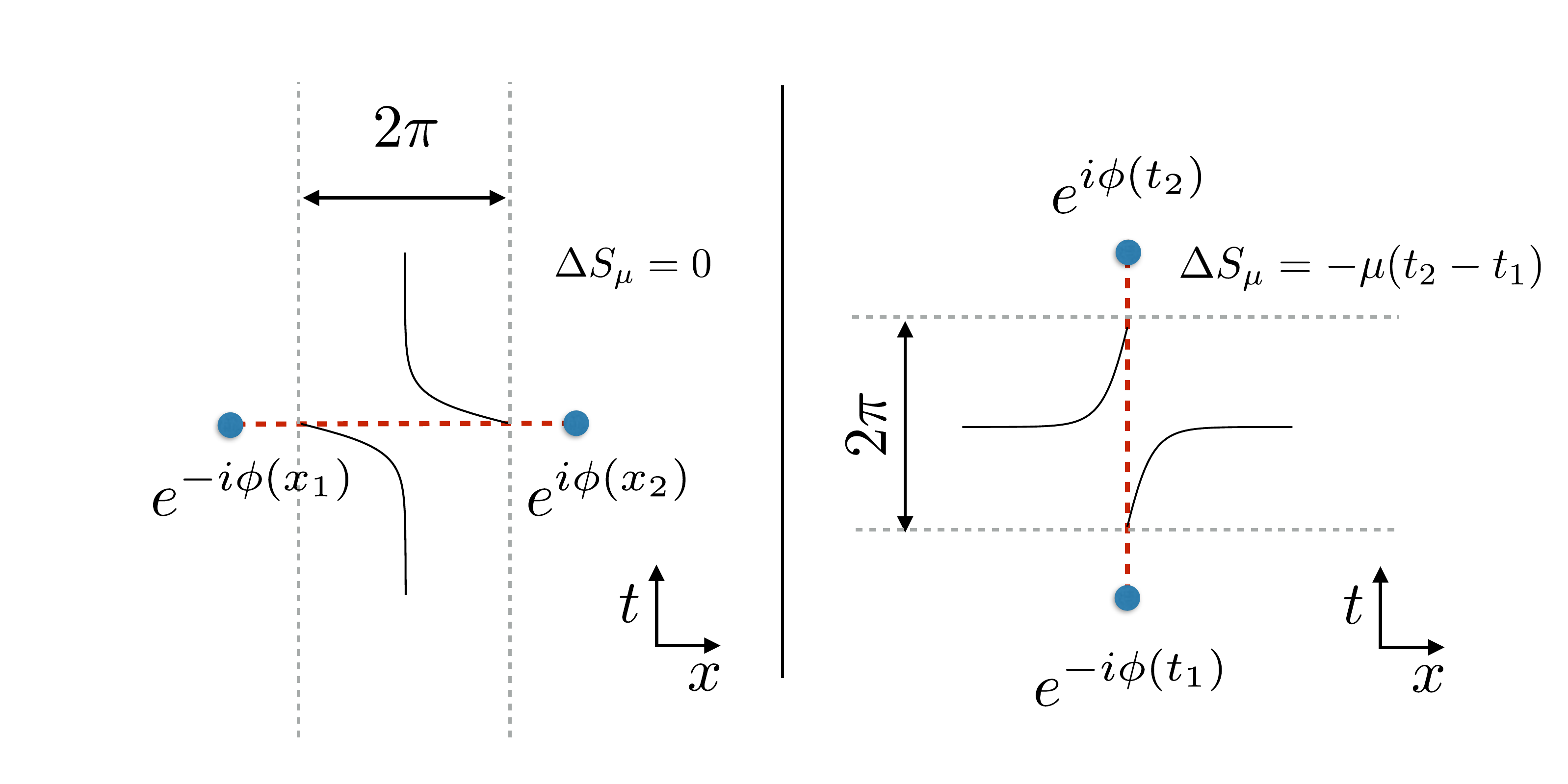} 
   \caption{The spatial (left) and temporal (right) correlators $\avg{e^{-i\phi({\bm x}_1)}e^{i\phi({\bm x}_2)}}$ in the $O(2)$ model. A jump in the $\sigma$-field must appear on a contour $C$ between the two sources, that can be chosen as a straight line. This  requires a kink solution (shown as a solid black curve) interpolating from the vacuum of the dual theory on both sides of the contour and having an appropriate $2\pi$ jump across $C$. Only the temporal correlator become dependent on the chemical potential.}
   \label{fig:O2confmu}
\end{figure*}

Without chemical potential the $O(2)$ model has two phases as a function of the bare coupling $g_0^2$, that are intimately related to vortices. At weak coupling, the two-point correlator decays algebraically, which is the maximal order in two dimensions, the $O(2)$ symmetry is still intact and vortices come in pairs. At strong enough coupling, a (Berezinsky-)Kosterlitz-Thouless \cite{Berezinsky:1970fr,Kosterlitz:1973xp} transition occurs  and the correlator decays exponentially. This disorder is caused by \textit{vortex condensation}. In the Sine-Gordon model it corresponds to a phase, where the potential term dominates and $\sigma$ is locked to one of its minima. For a quantitative analysis one has to study the renormalization group (RG) flow of the coupling and the fugacity \cite{Jose:1977gm,Nienhuis:1984wm,Boyanovsky:1988ge,Boyanovsky:1989mc,Boyanovsky:1990iw,Amit:1979ab} (see also \cite{Altland:2006si,Anber:2011gn} for simple, pedagogical derivations). Note that chemical potentials are not renormalized 
as they couple to conserved charges.

That the mass parameter $m$ is connected to the correlation length in the disordered phase can be understood easily. As discussed before, the dual field $\sigma$ has to jump by $2\pi$ across the contour $C$. Outside of $C$, it levels off to a vacuum value, see Fig.~\ref{fig:O2confmu}. The $\sigma$-profile perpendicular to $C$ is basically that of a kink solution cut in half. Thus, the field is concentrated in a strip of the width of a vortex and length $|{\bm x}_1-{\bm x}_2|$ (for the shortest contour). Its action is the kink mass, in our conventions $4mg_0/\pi\equiv M$, times this length. Hence, the two-point function \eqref{eq:two_point} behaves as $e^{-|{\bm x}_1-{\bm x}_2|M}$ generating an exponential decay/a linear interaction.

With chemical potential, the spatial and temporal correlators are different. For \textit{temporal} correlators the kink solution has its profile in the spatial direction (Fig.~\ref{fig:O2confmu}, right panel), the $x$-integration of $\partial_x\sigma/(2\pi)$ yields the kink charge $1$ in the range $t\in[t_1,t_2]$, and hence the action has an extra contribution $-\mu(t_2-t_1)$. For \textit{spatial} correlators, on the other hand, the kink is in the temporal direction (Fig.~\ref{fig:O2confmu}, left panel), $\partial_x\sigma$ vanishes and $\mu$ does not appear. 

Before discussing more effects of $\mu$, we want to stress the following point: The chemical potential, coupling only to a total derivative, does not affect the vortex interactions, and will therefore not affect statements about the KT transition.

Therefore, we now come to $\mu$-effects in the vortex condensed phase at strong coupling. The vacuum remains unchanged as long as $\mu<M$. Indeed, in this regime, the $\mu$-dependent part of the free energy can be approximated as the sum over kinks in the $x$-direction
\begin{equation}
\beta F=-\ln Z\sim-\ln\left[\sum_{Q,\bar Q}\frac{1}{Q!\bar Q!}e^{-M\beta (Q+\bar Q)+\mu \beta (Q-\bar Q)}\right]\;,
\end{equation}
where $Q$ and $\bar{Q}$ are the number of kinks and anti-kinks, respectively, and we have employed the dilute kink-anti-kink gas approximation, which is justified for low temperature (large $\beta$). The above looks like a partition function of indistinguishable classical particles with chemical potential. This expression, however, takes no account of the Bose-Einstein nature of kinks, for which we need to quantize their paths. To do this, let us endow the kink position ${\bm X}$ with (Euclidean) time dependence ${\bm X}(t)$, in which case the action of the kink becomes
\begin{multline}
 \int_0^\beta \!dt \!\int \!dx\; \Bigg\{\frac{g_0^2}{2(2\pi)^2}
 \Big[(\partial_t\sigma_{kink})^2+(\partial_x\sigma_{kink})^2\Big]\\-m^2\cos\sigma_{kink}-\mu\frac{\partial_x\sigma_{kink}}{2\pi}\Bigg\}=\\
 =\int_0^\beta dt\;\frac {M \dot {\bm X}(t)^2}{2}+(M-\mu)\,\beta\;,
\end{multline}
which is just an action of non-relativistic particles with ground state energy $M$ and chemical potential $\mu$ at temperature $\beta$. The anti-kink has a similar action except that $\mu\rightarrow -\mu$. Quantizing a system of particles with the above action produces a well known partition function of bosons, independent of $\mu$ at zero temperature ($\beta\to\infty$).  

So far we have ignored the interactions between kinks, which is justified as long as the kinks are well separated, i.e. if their densities are much larger than their witdth. When $\mu\ge M$, however, the kinks condense and a finite density of them is generated in the vacuum\footnote{Note, however, that kinks \emph{do not} Bose condense, as this is impossible for one-dimensional Bose gases, even with interactions.}. In this case one cannot ignore the interactions between kinks anymore as they become strongly overlapping and dense. Hence, semiclassical methods are not applicable. The qualitative picture of the correlators $\avg{e^{i\phi({\bm x}_1)}e^{-i\phi({\bm x}_2)}}$ should be as follows: these operator insertions can be interpreted as source at ${\bm x}_1$  and a sink at ${\bm x}_2$ for the conserved charge (see Sec. \ref{sec:lattice}). For $\mu>M$ the vacuum is filled by kink worldlines flowing in the time direction. Still the spatial correlators require the transfer of conserved charge in the spatial 
direction, presumably by some analogue of the spatial kink of the dilute phase. Since it does not couple to the chemical potential, its energy cost should still be proportional to the mass. It is therefore feasible that no significant change will occur in the spatial correlators\footnote{Note, however, that as the spatial 
separation of source and sink points reaches the density scale, some change in behavior should occur in the correlator. Namely at this point the conserved charge worldline connecting the source and the sink may break upon intersection with the condensed worldlines. Nevertheless the net transfer of conserved charge in space still has to occur, and, since worldlines are heavy, the main contribution to the action cost of this charge transfer will be $M|x_1-x_2|$.}.

\section{$O(3)$: chemical potential and phase diagram}
\label{sec:O3}

In the case of the $O(3)$ model, adding any chemical potential can be viewed as adding one for the rotations around an arbitrary component of $\bm n$, without loss of generality this has been taken to be the 3rd component\footnote{The conserved charge in $O(3)$ is the angular-momentum projection onto some axis.}. The Lagrangian is given by \eqref{eq:Lagain1} and contains a term proportional to $-\mu^2(n_1^2+n_2^2)=\mu^2 n_3^2$. This is an explicit mass term for the $n_3$-component, whose fluctuations are therefore suppressed across length scales larger than $1/\mu$. Thus, for large chemical potentials, the $O(3)$ theory abelianizes to $O(2)$. If we parametrize the field $\bm n$ with usual spherical coordinates $(\vartheta,\phi)$, the low energy effective theory is therefore described by the Lagrangian 
\be
\mathcal L_{\text{eff}}=\frac{1}{2g^2}\left((\partial_\nu\phi)^2+2i\mu\dot\phi-\mu^2\right)\;.
\ee
Several points about this Lagrangian are in order
\begin{itemize}

\item  The Lagrangian is valid for $\mu\gg \Lambda$, where $\Lambda$ is the strong scale of the theory.

\item It is identical to the $O(2)$ Lagrangian discussed in the previous section and therefore enjoys a dual description in terms of the $\sigma$ fields, cf.~\eqref{eq:action_O2_with}. In the light of the first point, this is an approximate dualization; in \sec{sec:lattice_O3} we give exact dualizations of the $O(3)$ model on the lattice.

\item Unlike the $O(2)$ model, the coupling $g^2$ is not a parameter of the model, but is the \emph{running coupling at scale $\mu$}, see below.

\item The full $O(3)$ theory has vortex-like \emph{meron} solutions carrying half-integer topological charge, much like those discussed in \cite{Affleck:1988nt}. Outside of their cores the vortices consist purely of ${\bm n}_\parallel$, the field space of the $O(2)$-model, in which they perform a winding. At their cores they visit the north or south pole of $O(3)$, $({\bm n}_\parallel,{\bm n}_\perp)=(0,0,\pm 1)$, respectively, without a UV-diverging action. Altogether there are four types of merons.

\end{itemize}

The occurrence of an $O(2)$-like phase for $\mu$ increasing from zero is related to a phase transition, that on general grounds is expected when $\mu$ reaches some $\mu_c$, the mass $M$ of the lowest excitation. This ends the so-called Silver Blaze phenomenon, which means the $\mu$-independence of the partition function and all its derivatives for $\mu\leq \mu_c$. In this phase transition the corresponding  charge condenses and the system abelianizes to an $O(2)$ model (selected through the explicit symmetry breaking in the choice of $\mu$). Hence, this $\mu_c$ basically sets a lower bound on $O(2)$-like effects to be discussed now\footnote{The anisotropies in  ${\bm n}$-space used in \cite{Haldane:1982rj,Affleck:1985jy,Vachaspati:2011ad} drive the system to an $O(2)$-like phase as well, but the Silver Blaze phenomenon seem not to apply.}. In fact, as in $O(2)$ and large $N$ expansion of $O(N)$ (see Sec. \ref{sec:largeN}), we expect that $\mu_c=M$, where $M$ is the 
mass gap of the order of the strong scale~$\Lambda$. 

Let us look at the Kosterlitz-Thouless argument -- which gives the exact critical coupling of vortex condensation in the $O(2)$ model -- for these $O(3)$ merons. Both action and entropy of the merons inherit a logarithmic divergence on the extent $L$ of the system from their vortex-like behavior outside of their cores. Their action, however, has a prefactor $1/g_{\text{eff}}^2$ and thus the existence of a KT transition at $\mu\geq\mu_c$ hinges on the \textit{strength of this coupling}. As we have already mentioned, $g_{\text{eff}}^2$ is the effective RG coupling at the scale given by $\mu$, i.e.\ when shells of momenta larger than $\mu$ have been integrated out in a Wilson RG manner.

The answer to this question seems clear for $\mu\gg \mu_c$: by one loop running of the coupling, $g^{-2}\sim \ln \frac{\mu}{\Lambda}$ (the exact dependence depends on the subtraction scheme), the effective coupling $g^2(\mu)$ is too weak for merons to percolate (their action suppression is too high to be compensated by the entropy). Then the spatial correlators are algebraic,
\be
\avg{\bm n(x_1)\bm n(x_2)}={|x_1-x_2|}^{-\frac{g^2_{\text{eff}}}{2\pi}}\;.
\ee

Nevertheless, it is feasible that there is a region of $\mu\gtrsim\mu_c$ where the coupling is strong enough for merons to percolate. In this phase the spatial correlators are exponentially decaying. The system would then undergo two transitions as $\mu$ is increased: the charge condensation transition at $\mu=\mu_c$ and a Kosterlitz-Thouless transition at some $\mu=\mu_{KT}>\mu_c$.

For the details of this running coupling, one has to combine the perturbative running in $O(3)$ mentioned above with the running in the abelianized theory of $O(2)$, where the coupling together with the meron fugacity runs nonperturbatively.

\begin{figure}[htbp] 
   \centering
   \includegraphics[width=.45\textwidth]{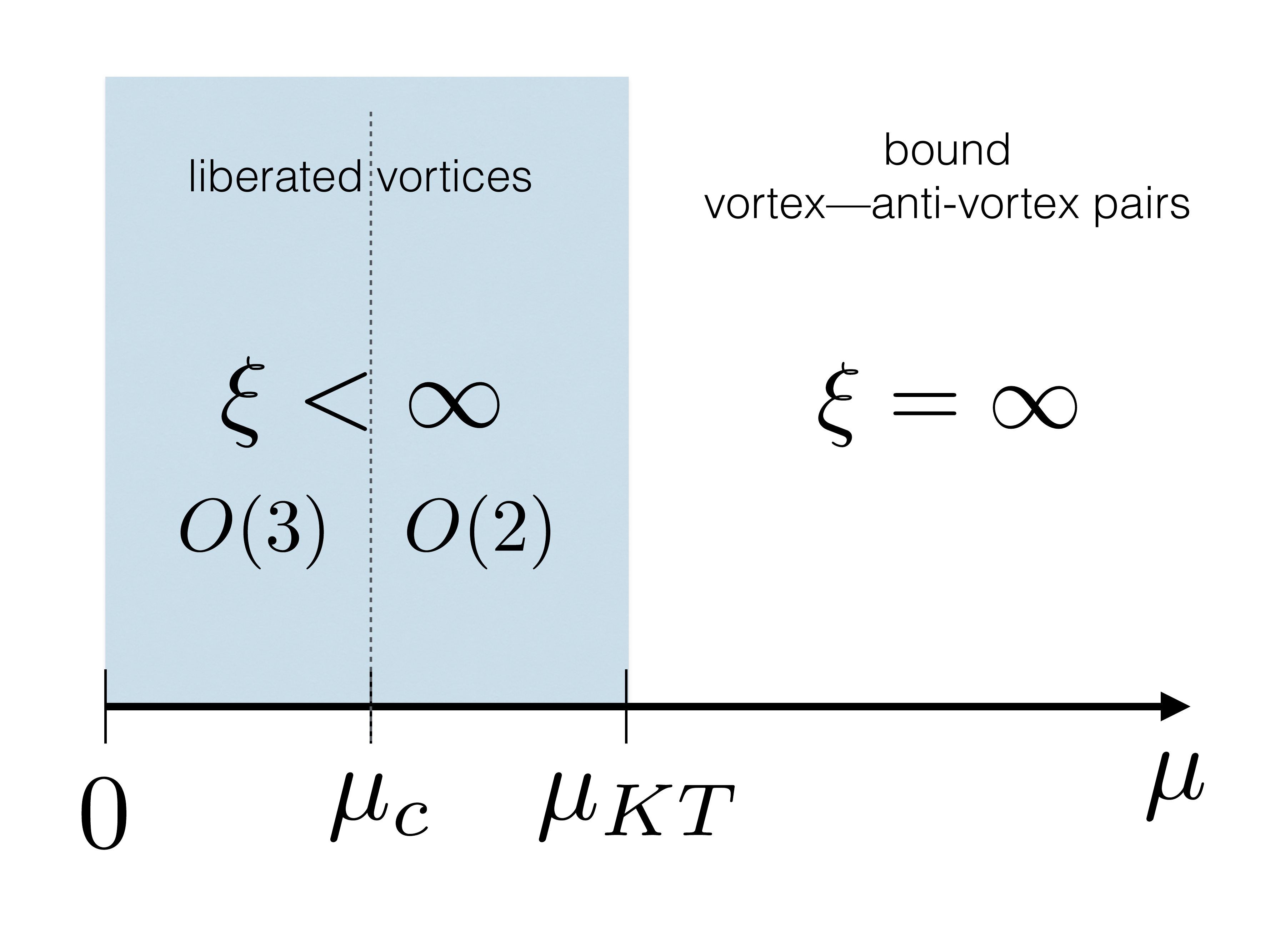} 
   \caption{The phase diagram of $O(3)$ as a function of a single chemical potential $\mu$. The shaded area represents the region where vortices (merons) condense and $\xi$ is the spatial correlation length.}
   \label{fig:phasediag}
\end{figure}

One must keep in mind, that it is possible that both transitions happen simultaneously. From the mass gap at $\mu=0$ the spatial correlators are expected to be exponentially decaying, which already implies a sufficient amount of disorder. If the KT transition was to occur exactly at $\mu_c$, we would expect that the spatial  correlation length diverges exponentially as the coupling approaches the critical value. In other words the inverse correlation length would go to zero. But as we already said, we expect that at $\mu\lesssim\mu_c=M$  the spatial correlation length should already be $1/M$. We therefore believe that the KT transition occurs at $\mu=\mu_{KT}>\mu_c$ and that as $\mu$ is lowered the density of merons increases sufficiently so as to exactly produce the spatial correlation length of $1/M$. If this is shown to be the case, it would be a very strong argument that merons are responsible for the spatial correlation length and -- since up to the $\mu$-dependent part we discussed in \sec{sec:O2}, the 
temporal correlators are subject to the same exponential decay -- for the mass gap generation in the $O(3)$ model. We should warn, however, that it is still possible that something entirely different happens at $\mu_c$ and that the inverse spatial correlation length experiences a discontinuous jump across $\mu_c$. Nevertheless, we will see that the large $N$ analysis supports our conjecture for the $O(3)$, and also that it indicates a second order phase transition at $\mu=\mu_c$.

In addition, because merons carry fractional topological cha\-rge, the KT phase transition would also be characterized by a change in the $\theta$-angle behavior of the partition function \cite{Affleck:1988nt}.
which would amount to replacing $m^2 \to m^2\cos(\theta/2)$ in the dual action. The change in the $\theta$-angle behavior would be from the meron-like $\cos(\theta/2)$ for $\mu<\mu_{KT}$ to the instanton-like $\cos\theta$ for $\mu>\mu_{KT}$. The percolation of merons should therefore be signalled by the jump in the topological susceptibility. Recall, however that in the condensed phase, the dual $\sigma$ field has no specific vacuum value, as it is required to change to accomodate kink condensation. Nevertheless the dependence on the $\theta$-angle can only enter as $\cos(\theta/2)$. On the other hand for $\mu>\mu_{KT}$ the theta dependence should be purely instanton like. Since the instantons are meron--anti-meron (of different type) bound pairs, they will not couple to the dual sigma field, and will therefore survive in the IR, turning the $\theta$-angle behavior to $\cos\theta$. We must stress, however, that small instantons have an UV divergent contribution, and would dominate the theta dependence in the 
continuum limit extrapolation. Nevertheless, for sufficiently coarse lattices we expect that the topological susceptibility measurement as a function of the chemical potential can be a good indicator of the KT transition. Alternatively one may try to employ the same treatment for higher $CP(N-1)$ models, where small instantons are not UV divergent and where the topological susceptibility has no aforementioned UV problems.

The situation described above is summarized in the phase diagram of Fig. \ref{fig:phasediag}. To give further support to the conjecture that $\mu_{KT}>\mu_c$ we will turn now to the large $N$ expansion of the $O(N)$ model.

\section{Large $N$ expansion}
\label{sec:largeN}

In this section we consider the large $N$ expansion of the $O(N)$ theory with chemical potential in the $(n_1,n_2)$-plane. From the Eqs.~(\ref{eq:Lfirst}-\ref{eq:n_decomp}) we write
\be
 S=\frac{N}{2\gth}\int \!d^2x\; 
 \Big\{\big[(\partial_\nu +\delta_{\nu 0}\mu\sigma_2){\bm n}_\parallel\big]^2
 +(\partial_\nu{\bm n}_\perp)^2\Big\}\;.
\ee
To ensure proper scaling with large $N$, we have also replaced $1/g_0^2=N/\gth$, so that now $\gth$ is the (bare) 't Hooft coupling. 

From $\mu=0$ it is well known that it is useful to eliminate the constraint ${\bm n}_\parallel^2+{\bm n}_\perp^2=1$ by the introduction of a Lagrange multiplier\footnote{The contour of integration over $\lambda$ is along (or parallel to) the real axis in the complex $\lambda$ plane.} $\lambda$. The action becomes quadratic in the ${\bm n}$-fields,
\begin{align}
 S=&\,\frac{N}{2\gth}\int \!d^2x\;
 \Big\{{\bm n}_\parallel \Gamma_\parallel {\bm n}_\parallel+
 {\bm n}_\perp \Gamma_\perp {\bm n}_\perp-i\frac{\gth}{N}\lambda\Big\}\\
 &\Gamma_\parallel=-\big(\partial_\nu 1_2 +\delta_{\nu 0}\mu \tau_2\big)^2+i\frac{\gth}{N}\lambda\, 1_2\\
 &\Gamma_\perp=\big(-\partial_\nu^2+i\frac{\gth}{N}\lambda\big)\,1_{N-2}\;,
\end{align}
where we have explicitly written out that the $\Gamma$'s act on $2$ respectively $N-2$ components. 
The Gaussian integration over the $\bm n$-fields can now be performed and yields the effective action
\be\label{eq:LargeNeffaction}
 S_{\text{eff}}
 =\frac{1}{2}\Tr\log\Gamma_\parallel
 +\frac{1}{2}\Tr\log\Gamma_\perp
 -\frac{N}{2\gth}\int\! d^2x\;i\frac{\gth}{N}\lambda\;,
\ee
where we have neglected a term $\Tr\log(N/\gth)$ because it does not depend on the fields nor $\mu$. 
In principle, this action\footnote{A similar action for a \textit{linear} sigma model in $3+1$ dimensions has been used in \cite{Andersen:2006ys} in the context of pion condensation at finite isospin density.} is nonlocal (has infinitely many local terms). 

For large $N$, however, one can evoke a saddle point approximation in complete analogy to the $\mu=0$ case. Note that since $\Gamma_\perp$ acts on $N-2$ components, the middle term in \eqref{eq:LargeNeffaction} belongs to the leading order in $N$, whereas the first term is subleading\footnote{Choosing the chemical potential to act on more components, this term can be made leading order in $N$, too.}. It follows immediately, that the saddle point to leading order is $\mu$-independent and so is the dynamical mass derived from it. The saddle point is taken to be an (imaginary) space-time independent vacuum expectation value of $\lambda$ and the fluctuations $\dlam$ of the field around it are suppressed by $\sqrt{N}$,
\be
 i\,\frac{\gth}{N}\lambda=m^2+\frac{i}{\sqrt{N}}\,\dlam(x)\;.
\ee
The saddle point equation to leading order,
\be
 \frac{N-2}{2}\int\frac{d^2k}{(2\pi)^2}\frac{1}{k^2+m^2}
 =\frac{N}{2\gth}\;,
\label{eq:saddle_leading}
\ee
yields, again to leading order, the dynamical mass $m^2=\Lambda_{\text{UV}}^2\,e^{-4\pi/\gth}$ with $\Lambda_{\text{UV}}$ the ultraviolet cut-off of the momentum integral (or the Pauli-Villars mass). Asymptotic freedom in the running coupling of the $O(N)$ models ensures that this combination of cut-off and bare coupling is a renormalization group invariant, namely the scale $\Lambda$ of the model, $m^2=\Lambda^2$, which holds even for finite $N$.

We will now study the effects of the chemical potential by plugging the mass and the fluctuations $\dlam$ back into the Lagrangian. Like at $\mu=0$ it is convenient to reparametrize the $\Tr \log$s including the mass term by writing
\begin{align}
 \mathcal L
 =\frac{N}{2\gth}\int \!d^2x\;
 \Big\{&{\bm n}_\parallel \big[-(\partial_\nu +\mu\delta_{\nu 0} \tau_2)^2+m^2\big] {\bm n}_\parallel\nonumber\\
 +&{\bm n}_\perp \big[-\partial_\nu^2+m^2\big] {\bm n}_\perp\\
 +&\frac{i}{\sqrt{N}}\,\dlam ({\bm n}_\parallel^2+{\bm n}_\perp^2)
 -(m^2+\frac{i}{\sqrt{N}}\,\dlam)\Big\}\nonumber\;.
\end{align}
Both fields ${\bm n}_\parallel$ and ${\bm n}_\perp$ interact with $\dlam$ through $\sqrt{N}$-suppressed three-vertices and the corresponding perturbation theory could be used now. 

The chemical potential only enters in the propagator of ${\bm n}_\parallel$, with momentum $q_\nu$ the latter is $1/((q_\nu\pm i \delta_{\nu 0}\mu)^2+m^2)$. For small chemical potentials, $\mu<m$, the partition function is $\mu$-independent: it is made of diagrams without external legs and its $\mu$-dependence can only come from virtual ${\bm n}_\parallel$-lines. Every ${\bm n}_\parallel$-line can branch off $\dlam$-lines, say $K$ of them, and eventually needs to close. 
The corresponding term in the partition function contains $K+1$ propagators
\begin{align}
 &\,\frac{1}{(q_\nu\pm i \delta_{\nu 0}\mu)^2+m^2}\,
 \frac{1}{(q_\nu+p_\nu^{(1)}\pm i\delta_{\nu 0}\mu)^2+m^2}\times\ldots\nonumber\\
 &\,\times \frac{1}{(q_\nu+p_\nu^{(K)}\pm i \delta_{\nu 0}\mu)^2+m^2}\;,
 \label{eq:prop_parallel}
\end{align}
where we used that $\tau_2$ can be diagonalized with eigenvalues\footnote{This eigenvalue is the same in all propagators since it is conserved at the vertices.} $\pm 1$. Specifying this sign to be $+1$ for a moment, the poles in the complex $q_0$-plane are at $-p_0^{(i)}-i\left(\mu\pm\sqrt{(q_1+p_1^{(i)})^2+m^2}\right)$. Hence the integration contour in $q_0$ can be shifted from $\mathbb{R}$ to $\mathbb{R}-i\mu$ (or to $\mathbb{R}+i\mu$ for the other sign) without crossing any poles and the momentum integration is independent of $\mu$. This can be done for all ${\bm n}_\parallel$-loops separately. Therefore, the entire partition function is independent of $\mu$, which is again a manifestation of the Silver-Blaze phenomenon. Since the charge expectation value is just the $\mu$-derivative of $\ln Z$, it vanishes and no charged state is excited in this regime.

As the fields ${\bm n}_\perp$ carry no $\mu$-dependence and appear quadratically in the Lagrangian, they should be integrated out. At the same time this will yield the leading order contribution, $N-2$ fields out of $N$, to the effects discussed below. In deriving an effective Lagrangian one integrates out the high momenta of ${\bm n}_\parallel$ as well. For large chemical potentials $\mu\geq m$ the contribution of ${\bm n}_\parallel$ to the e.g.\ the saddle point equation is 
\be
 \int\frac{d^2q}{(2\pi)^2}\,\frac{1}{(q_\nu\pm i \delta_{\nu 0}\mu)^2+m^2}
 =\!\int\limits_{q_1^2\geq\, \mu^2-m^2}\!\!\frac{dq_1}{4\pi}\,\frac{1}{\sqrt{q_1^2+m^2}}\;.
\ee
Only momenta $q_1^2\geq\, \mu^2-m^2$ contribute. The tachyonic behaviour in the complementary regime of small momenta is another sign of the expected charge condensation. For high momenta, $q^2\geq\mu^2$, one can evoke the argument around \eqref{eq:prop_parallel} to see that this contribution is $\mu$-independent. It thus agrees with that of ${\bm n}_\perp$ and all $N$ fields ${\bm n}$ contribute in an $O(N)$ invariant manner, like at vanishing chemical potential.

The interaction with the propagating fields ${\bm n}$ provides a two-point function for $\dlam$, again this calculations proceeds like for the $\mu=0$ case,
\begin{align}
 \!\!S\propto&\,\frac{1}{2}\int d^2k\,\dlam(k)\Gamma(k)\dlam(-k)\\
 \!\!\Gamma(k)=&\,\int\!\frac{d^2q}{(2\pi)^2}\frac{1}{q^2+m^2}\frac{1}{(k-q)^2+m^2}
 =\frac{1}{4\pi m^2}+\mathcal{O}(k^2)\;,
\end{align}
up to $1/N$ corrections. The exact result for $\Gamma$ can be found in \cite{D'Adda:1978uc}, but is not very illuminating (apart from the fact that $\dlam$ has no particle excitation in the usual sense).

Our main point will be that the interaction with $\dlam$ induces an ${\tilde{\bm n}}_\parallel^4$-term in the low energy effective action, where $\tilde{\bm n}$ is the low-momentum part of $\bm n$ which we keep in the low energy effective action. This effect is very analogous to a four-fermi interaction induced (at low energies) by an intermediate heavy boson, that interacts with the fermions through a three (Yukawa-)vertex and whose $\Gamma$ is $\text{mass}^2+\mathcal{O}(k^2)$. To be precise, the effective action contains the following terms
\begin{align}
 S_{\text{eff}}
 \propto\frac{1}{2}\int \!d^2k \Big\{\dlam(k)\Gamma(k)\dlam(-k)+i\frac{\sqrt{N}}{\gth}\dlam(k)\tilde{\bm n}_\parallel(-k)\Big\}\;,
\end{align}
and the Gaussian integration over $\dlam$ yields
\begin{align}
 S_{\text{eff}}
 \propto\frac{1}{2}\int\!d^2k\, \tilde{\bm n}_\parallel^2(k)\Big(\frac{\pi N m^2}{2\gthsq}+\mathcal{O}(k^2)\Big)\tilde{\bm n}_\parallel^2(-k)\;.
\end{align}
Collecting the ${\bm n}_\parallel$ terms in the effective action, 
\begin{align}
 \mathcal{L}_{\text{eff}}
 \propto&\, \frac{N}{2\gth}\int \!d^2x\; \Big\{
 \tilde{\bm n}_\parallel \big[-(\partial_\nu +\delta_{\nu 0}\mu \tau_2)^2
 \big] \tilde{\bm n}_\parallel 
 +m^2\tilde{\bm n}_\parallel^2
 +\lampar \tilde{\bm n}_\parallel^4\Big\}\nonumber\\
 &\lampar=\frac{\pi m^2}{\gth}\;,
 \label{eq:Leff_nparallel}
\end{align}
we see that the $\tilde{\bm n}_\parallel^4$ term is of the same order in $N$ as the others. 
This system is nothing but a massive scalar (`Higgs') field with quartic interaction subject to a chemical potential. From here on we can take over textbook results, albeit in two dimensions. 

For large chemical potentials, $\mu>m$, an instability of the quadratic term $(-\mu^2+m^2)\tilde{\bm n}_\parallel^2$ occurs (that invalidates the treatment of free scalars), but is acted against by the quartic term $\lampar \tilde{\bm n}_\parallel^4$ in \eqref{eq:Leff_nparallel}.  
A nontrivial minimum develops at constant $\tilde{\bm n}_\parallel^2=(\mu^2-m^2)/(2\lampar)$, i.e.\
\be
\avg{\tilde{\bm n}_\parallel^2}=\frac{\gth(\mu+m)}{2\pi m^2}\,(\mu-m)\qquad (\mu\geq m)\;.
\ee
At its minimum, the potential is $-N(\mu^2-m^2)^2/(8\gth\lampar)$. The expectation value of the charge density is minus the $\mu$-derivative of that value,
\begin{align}
 \frac{1}{V}\avg{q}
 =&\,\frac{N\mu}{2f_0^2\lampar}(\mu^2-m^2)+\ldots\nonumber\\
 =&\,\frac{N\mu(\mu+m)}{2\pi m^2}\,(\mu-m)+\ldots\quad (\mu\geq m)
\end{align}
and thus increases linearly at $\mu=m$. The ellipses above indicate the contribution from the imaginary part of the action.
This means a second order transition to a phase where the symmetry is broken to $O(2)$. Notice that the charge is given in terms of the mass gap $m$, not the unphysical bare charge $\gth$ in the UV.

We have discussed action minima for constant fields in mean field theory. 
More detailed studies of this system (at finite $N$ and in four dimensions) have been done through complex Langevin and dual variable simulations \cite{Aarts:2009hn,Gattringer:2012ap,Gattringer:2012df}. They confirmed the Silver-Blaze phenomenon ending in a second order transition including an expecation value of the squared Higgs field. The critical chemical potential was found to be  shifted from the free expectation by the interactions and the effect of $\mu$ on the temporal correlators discussed in Sec.~\ref{sec:O2} was seen as well. 

To summarize, we have performed a large $N$ approximation along the lines of the $\mu=0$ calculation, resulting in a dynamical mass. The ${\bm n}_\parallel$-fields also interact with the fluctuating part of the Lagrange multiplier field $\dlam$, which is to be integrated out, too\footnote{therefore the effects coming from this can presumably not be seen in linear sigma models}, giving rise to an effective $\tilde{\bm n}_\parallel^4$-theory at nonzero density. The Silver-Blaze phenomenon is expected to end near the mass gap giving both the charge and the $\tilde{\bm n}_\parallel^2$ field nonzero expectation values, thereby breaking the symmetry down to $O(2)$. The latter is a prerequisite for the Kosterlitz-Thouless transition at larger chemical potentials.

\section{Lattice formulations}
\label{sec:lattice}

\subsection{$O(2)$}
\label{sec:lattice_O2}

We first consider the continuum Lagrangian (\ref{eq:Lagain1}), which in terms of the complex field $n=n_1+in_2$ reads,
\begin{align}
 \g^2\mathcal L
 =-\frac{1}{2}n^*(\partial_\nu-\delta_{\nu 0}\mu)^2n\;.
 \label{eq:L_O2_compl}
\end{align}
In the lattice discretization one works with exponential factors $\exp(\mu a)$ with $a$ the lattice spacing. To be precise, let $\hat{\nu}$ the unit vector in the $\nu$-direction and $x=l_x a$ and $y=l_y a$ lattice sites with $l_{x,y}$ two-vectors with integer components. A possible lattice action for the $O(2)$ model with chemical potential is 
\begin{align}
 S\propto-\frac{1}{2}\sum_{x,y}n^*(x)\cdot\sum_\nu M_\nu(x|y)\cdot n(y)\;,
 \label{eq:lattice_O2_action}
\end{align}
with the matrices
\begin{align}
 M_\nu(x|y)=e^{-\mu a\delta_{\nu 0}}\delta_{x+\hat{\nu}a,y}
 +e^{\mu a\delta_{\nu 0}}\delta_{x-\hat{\nu}a,y}
 -2\delta_{x,y}\;,
 \label{eq:lattice_O2_matrix}
\end{align}
where we have abbreviated $\sum_x\equiv \sum_{l_x}$ and  $\delta_{x,y}\equiv\delta_{l_x,l_y}$.
In the limit $a\to 0$ this action becomes the continuum action. 

Plugging in the parametrization $n=\exp(i\phi)$, shifting $x$ under the sum once and neglecting a constant one arrives at
\begin{align}
  S=-J\sum_{x,\nu}\cos\big(v_\nu(x)-i\delta_{\nu0}\mu a\big)\;,
 \label{eq:O2_lattice_naive}
\end{align}
where $J$ is a dimensionless parameter replacing $1/\g^2$ and we have introduced the (forward) lattice derivative of the field,
\begin{align}
 v_\nu(x)=\nabla_{\!\nu}\,
 \phi(x)\equiv\phi(x+\hat{\nu}a)-\phi(x)\;,
 \label{eq:O2_lattice_velocity}
\end{align}
in analogy to the velocitiy field in \sec{sec:O2}.
This action\footnote{This system has recently been studied for modelling optical lattices in the context of quantum computing \cite{Zou:2014rha}.} is manifestly periodic in the variable $\phi$ and at $\mu=0$ reduces to the $XY$ model. For the latter, a strong coupling expansion in small $J<2/\pi$ shows that the two point correlator of $e^{i\phi}$, \eqref{eq:two_point}, is obtained by the minimal number of insertions of $J\cos(v_\nu(x))$ along the line connecting $x$ and $y$ and thus decays exponentially with the distance. 
Once more, this action is not real at nonzero real chemical potentials.

An alternative action \cite{Jose:1977gm} can be obtained using $\exp(J\cos\Phi))=\sum_{k=-\infty}^\infty I_k(J)\exp(ik\Phi)$ with $I_k$ the modified Bessel functions, on each link, i.e.\ with integer link-variables $k_\nu(x)$. From the action (\ref{eq:O2_lattice_naive}) one arrives at the partition function

\begin{align}
 Z=\prod_{x,\nu}\sum_{k_\nu=-\infty}^\infty
 \delta\,\Big(\sum_\rho\nabla_\rho k_\rho\Big)\,
 e^{k_0\mu a}I_{k_\nu}(J)
 \qquad\big(k_\nu\equiv k_\nu(x)\big)\;,
\end{align}
where the delta function has emerged upon integrating out the original field $\phi(x)$ and imposes the conservation of the currents $k_\nu(x)$ at each site $x$. Upon shifting the $x$-argument, the constraint becomes
\be
 \sum_\nu\big[k_\nu(x)-k_\nu(x-\hat{\nu}a)\big]=0\;,
 \label{eq:currentcons}
\ee
where one recognizes the vanishing divergence of the integer valued vector field $k_\nu$. As expected, the lattice chemical potential $\mu a$ enters with the net number of temporal occupation numbers $k_0(x)$. An obvious advantage of this action is its positivity. Therefore, it is used for numerical simulations, under the name of `dual variables'\footnote{In numerical simulations the challenge is to update such a constrained system, for which worm algorithms are suitable, for a review see e.g. \cite{Gattringer:2014cpa}.}, but can also be modified further analytically. 

The conservation of the current $k_\nu(x)$ implies that it is the (two-dimensional) curl of a scalar, again integer-valued field~$m$: $k_\nu(x)=\epsilon_{\nu\rho}\nabla_{\!\rho}\, m(x)$. 
At weak coupling, i.e.\ large $J$, one can further approximate the Bessel functions as exponentials $I_k(J)\approx\frac{\exp(J)}{\sqrt{2\pi J}}\cdot\exp(-k^2/(2J))$ leading to the Villain action \cite{Villain:1974ir} (where the $k$-independent factor is an irrelevant multiplicative constant). After that the path integral can be cast back into the usual form with an action 
\begin{align}
 S=\sum_{x,\nu}\Big\{\frac{1}{2J}\,(\nabla_{\!\nu}\,m)^2+\mu\, a\, \nabla_{\!1}m\Big\}\;.
 \label{eq:action_in_scalars}
\end{align}
This action has been written down in \cite{Meisinger:2013zfa}. Note the similarity to the continuum action (\ref{eq:action_O2_without}) including the coupling of $\mu$ to the kink charge. By means of a Poisson resummation, the $m$-field can be traded for a continuous field $\sigma$ including a $\cos\sigma$ term. 

At strong coupling, i.e.\ small $J$, the Bessel function can be approximated by $I_k(J)\approx (J/2)^{|k|}/|k|!$ suppressing all  currents $k$. The chemical potential gives larger  weights to positive currents (for positive $\mu$s\footnote{Negative $\mu$s of course have the opposite effect.}) in the temporal direction, since all $|k_0(x)|$ in the exponent are now multiplied by $\sign(k_0(x))\mu a +\log\frac{J}{2}$. Together with the charge conservation, the most dominant $\mu$-dependent contributions are those with unit $k_0$ on straight temporal loops closed through periodicity. Their weights are 
\be
\exp\Big(\big(\mu+\frac{1}{a}\log\frac{J}{2}\big)L_t\Big)\;,
\ee
where the $N_t a=L_t$ is the physical extent in the temporal direction. At small $\mu$, the weight of such worldlines is still exponentially suppressed with $L_t$. For $\mu$s at the order of\footnote{For the precise value of this critical $\mu$ one has in addition to consider entropy factors, worldlines with detours etc.} $1/a\cdot\log J/2$ the worldlines start to emerge and the partition function will become $\mu$-dependent, indicating the end of the Silver blaze region within the $O(2)$ model. 

Furthermore, temporal currents will not influence the spatial correlators orthogonal to them that remain exponentially decaying, which will only change at the Kosterlitz-Thouless transition at even larger $\mu$.

For $O(2)$ we have stuck to the lattice formulation, while for higher sigma models asymptotic freedom allows to always consider the continuum limit.

\subsection{$O(3)$}
\label{sec:lattice_O3}

One generalisation to the $O(3)$ model with spherical coordinates $n_3=\cos\vartheta$, $n_1+in_2=\sin\vartheta e^{i\phi}$ and chemical potential in the $(n_1,n_2)$-components can be done in close analogy to (\ref{eq:lattice_O2_action}, \ref{eq:lattice_O2_matrix}),
\begin{align}
S\propto  -\frac{1}{2}\sum_{x,y}\Big\{(\sin\vartheta\, e^{-i\phi})(x)
 \cdot&\sum_\nu M_\nu(x|y)\cdot 
 (\sin\vartheta\, e^{i\phi})(y)\nonumber\\
 +\cos\vartheta(x)\cdot&\sum_\nu M_\nu(x|y)\big|_{\mu=0}\cdot \cos\vartheta(y)\Big\}\;,
\end{align}
giving
\begin{align}
 S=-J\sum_{x,\nu}\Big\{&\cos\vartheta(x)\cos\vartheta(x+\hat{\nu}a)
 \label{eq:lattice_O3_action}\\
 +&\sin\vartheta(x)\sin\vartheta(x+\hat{\nu}a)
 \cos\big(v_\nu(x)-i\delta_{\nu0}\mu a\big)\Big\}\;,\nonumber
\end{align}
or, after introducing the currents $k_\nu(x)$ (abbreviated as $k_\nu$ again):
\begin{align}
 &Z=\prod_{x}\int\! d(\cos\vartheta)(x)
 \prod_{\nu}
 e^{J\cos\vartheta(x)\cos\vartheta(x+\hat{\nu}a)}\\
 &\times \sum_{k_\nu=-\infty}^\infty\!
 \delta\Big(\sum_\rho\nabla_\rho k_\rho\Big)\,
 e^{k_0\mu a}I_{k_\nu}(J\sin\vartheta(x)\sin\vartheta(x+\hat{\nu}a))\;.\nonumber
\end{align}
The appearance of the angle $\vartheta$ in comparison to the $O(2)$ case does not change the positivity of the path integral weight. Note that the variables of integration/summation are of different type: continuous angles\footnote{With more such angles, similar actions can be written down for $O(N)$-models.} $\vartheta(x)\in[0,\pi]$ on sites and integers $k_\nu(x)$ on links.

Let us discuss another dualization of the $O(3)$ model in close analogy to the $O(2)$ that yields integer variables only. For the moment consider the case $\mu=0$, see also \cite{Liu:2013nsa}. The lattice action is given by
\be
S=J\sum_{x,\nu}\bm n({x})\cdot\bm n({x+\hat\nu a})\;.
\ee
Since the 3-vectors are normalized $\bm n(x)^2=1$, the product reduces to a cosine, $\bm n(x)\cdot\bm n(x+\hat\nu a)=\cos\alpha_{\nu}(x)$, where $\alpha_{\nu}(x)$ is the angle between the $\bm n$-vectors at neighboring sites. By virtue of an expansion in Legendre polynomials we can rewrite the action in terms of the local spherical coordinates $(\vartheta(x),\phi(x))$ as follows
\begin{align}
 e^{J \cos\alpha_{\nu}(x)}
 =&\,\sum_{l_{\nu}=0}^\infty 
 c_{l_{\nu}}(J)\,P_{l_{\nu}}\big(\cos\alpha_\nu(x)\big)
 \label{eq:rewrite_exp(cos(alpha))}\\
 =&\,\sum_{l_{\nu}}c_{l_{\nu}}(J)\frac{4\pi}{2l_{\nu}+1}\,\times
 \label{eq:rewrite_Legendre}\\
 &\times\!\!\sum_{m_{\nu}=-l_{\nu}}^{l_{\nu}}\!\!
 Y^*_{l_{\nu}m_{\nu}}(\vartheta({x}),\phi(x))\,
 Y_{l_{\nu}m_{\nu}}(\vartheta(x+\hat \nu a),\phi(x+\hat\nu a))\;,\nonumber
\end{align}
where we have abbreviated $l_\nu\equiv l_\nu(x)$, $m_\nu\equiv m_\nu(x)$ and in the second line we have used the addition theorem for spherical harmonics. The constants $c$ can be computed from the orthonormality of the Legendre polynomials, for large $J$: 
\be
 c_l(J)\sim \frac{2l+1}{2}\frac{e^{J}}{J}e^{-\frac{l(l+1)}{2J}}\;.
 \label{eq:c_largeJ}
\ee

\begin{figure}[htbp] 
   \centering
   \includegraphics[width=.45\textwidth]{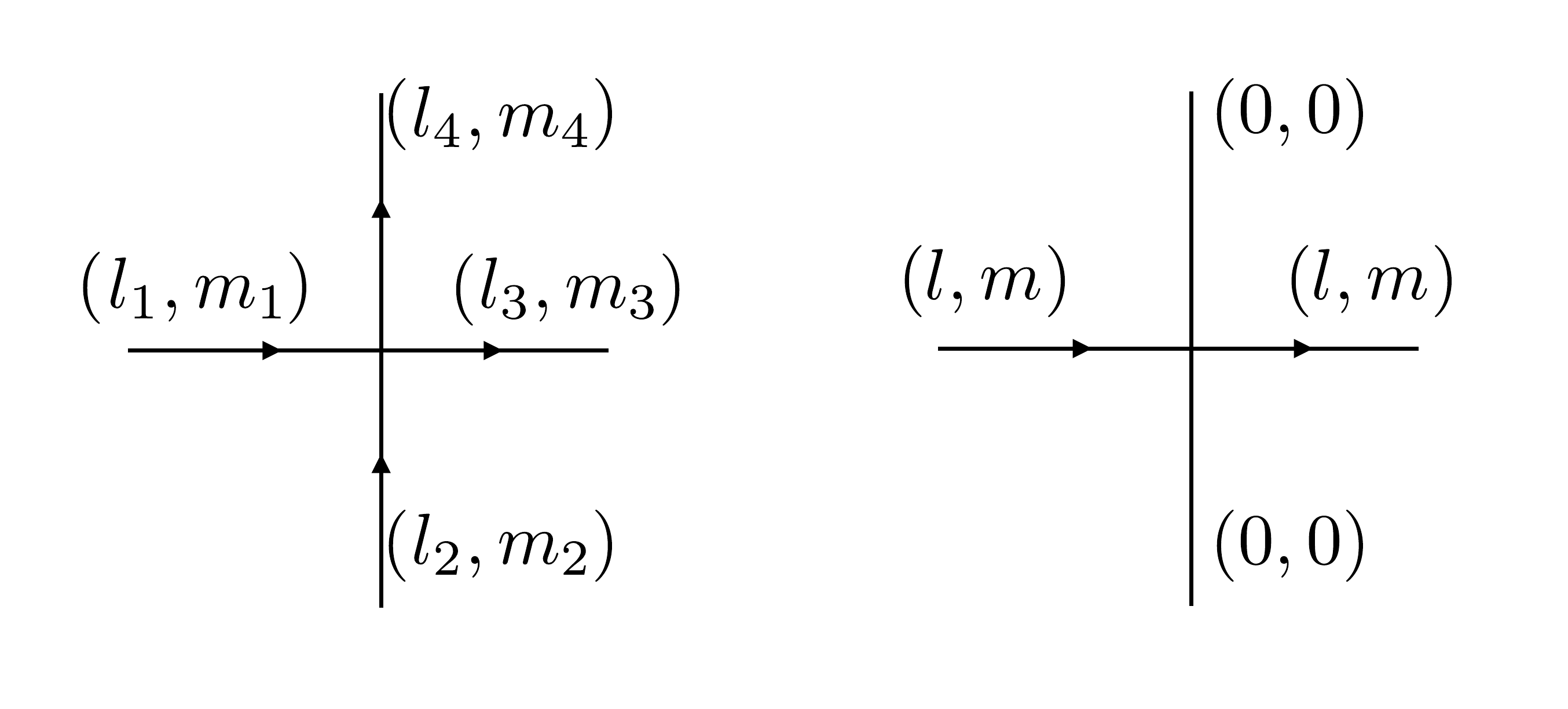} 
   \caption{A diagramatic representation of current conservation. The relation between the $m$-quantum number is $m_1+m_2=m_3+m_4$, while the incident $l$ quantum numbers obey the  \eqref{eq:interaction}. On the right, a current flow in the vacuum is shown, in which case both $l$ and $m$ is conserved. }
   \label{fig:currentcons}
\end{figure}

To obtain a dual theory in terms of the $l$ and $m$ integers, which are again link-variables, we must integrate over $(\vartheta(x),\phi(x))$. Since at each site $x$ there will be four spherical harmonics, we can define vertex coefficients
\be\label{eq:interaction}
 F_{\{l\},\{m\}}(x)
 =\!\int\! d\Omega_{x}
 Y^*_{l_{0}(x)m_{0}(x)}
 Y^*_{l_{1}(x)m_{1}(x)}
 Y_{l_{-0}(x)m_{-0}(x)}
 Y_{l_{-1}(x)m_{-1}(x)}\;,
\ee 
where we have dropped the argument $(\vartheta(x),\phi(x))$ and defined negative directions through $l_{-\nu}(x)\equiv l_\nu(x-\hat{\nu}a)$ etc. These $F$-terms represent the interactions of four currents incident on the site $x$ carrying the angular momentum quantum numbers $l$ and $m$, which are constrained to preserve angular momentum. For the $m$s this simply means 
the aforementioned constraint (\ref{eq:currentcons}), $\sum_\nu\big[m_\nu(x)-m_\nu(x-\hat{\nu}a)\big]=0$, for all $x$, while the $l$s are slightly more involved  (see below and Fig.~\ref{fig:currentcons}).

To add the chemical potential in the dual theory is simply done by replacing the difference\footnote{This is the way the product $Y^*_{.,m_\nu(x)}(.,\phi(x))Y_{.,m_\nu(x)}(.,\phi(x+\hat{\nu}a))$ in \eq{eq:rewrite_Legendre} for $\nu=0$ depends on the angles $\phi$.} $\phi(x+\hat 0 a)-\phi(x)\rightarrow \phi(x+\hat 0 a)-\phi(x)+i\mu a$, which introduces a term $e^{m_{0}(x)\mu a}$ for each temporal link (as in the previous dual actions).

Although no sign problem appears due to $\mu$, the $F$-terms may be positive or negative. It is not straightforward to see whether the product of all $F$-terms is of positive sign for individual loops. It does seem plausible, however, that negative $F$-terms always appear in even number of times in each loop, which is what happens for simple cases we examined. If this can be shown for any loop configuration, it would allow simulating the $O(3)$ model in this representation even at nonzero chemical potential.

Let us pause for the moment and see what the above $F$-terms mean. In the case of the $XY$ model, there was only $O(2)$ symmetry and, therefore, only the $m$ quantum number. So an expression equivalent to $F$ imposes the conservation of the $m_\nu(x)$ mentioned above. 

Similarily, consider a situation of an $O(3)$-link carrying quantum numbers $l,m$ and incident onto a point where no angular momentum flows in from two other links (see the right panel of Fig.~\ref{fig:currentcons}). It follows that the current goes through undisturbed. Incidentally this is precisely what happens in the case of a 3D rigid rotor, where only the time direction exists. The partition function of a quantum rotor reduces to
\be
 Z_{\text{3D-rotor}}=\sum_{l=0}^\infty\sum_{m=-l}^l\left(c_l(J)\frac{4\pi}{2l+1}\right)^{N_t}\;,
\ee
where again $N_t$ is the number of lattice sites in the time direction. 
For large $J$ (continuum limit), upon using \eqref{eq:c_largeJ}, we get the well-known partition function of the 3D rotor (up to an irrelevant constant)
\be
Z_{\text{3D-rotor}}=\sum_{l=0}^\infty\,(2l+1) \,e^{-\frac{l(l+1)}{2J}L_t}\;,
\ee
with $L_t$ the physical extent of the time direction.

In two dimensions the situation is more involved, because currents interact with other currents in a nontrivial way, governed by the $F$-term interaction \eqref{eq:interaction}. This interaction reflects the conservation of momentum:  the total angular momentum of the incoming currents equals the total angular momentum of the outgoing currents. Both can be computed through expressing the product of two spherical harmonics by a third one and the Clebsch-Gordon coefficients\footnote{Actually, the expression for $F$ can be seen as a generalisation of Wigner's $3j$-symbols (where three $Y$'s are integrated).}. This procedure is straightforward, and although a closed formula could be cumbersome, for a given set $\{l\},\{m\}$ (e.g.\ encountered in a simulation) $F$ can be computed easily.

Each link is additionally weighted by the $l$-dependent $c_l(J)$, which for large $J$ decreases like $e^{-l(l+1)/(2J)}$, \eqref{eq:c_largeJ}. Note that this is an analogue of the Villain limit of the $XY$ model, cf.\ above \eqref{eq:action_in_scalars}. 
In the Villain limit, one uses this action at small $J$, i.e.\ strong coupling, too. Now links with nonzero angular momentum $l>0$ become suppressed (just like nonzero $k$'s in $O(2)$), the currents are heavy.
In the two-point function the source and the sink creates and annihilates currents of angular momentum $l=1$ (and same $m\in\{-1,0,1\}$),
\be
 \bm n(x)\cdot\bm n(y)
 =\frac{4\pi}{3}\sum_{m=-1}^1 
 Y^*_{1m}(..(x))\, Y_{1m}(..(y))\;,
\ee
which follows from Eqs.~(\ref{eq:rewrite_exp(cos(alpha))},\ref{eq:rewrite_Legendre}).
As in the $O(2)$ case they have to be matched by dynamical currents and the cheapest way to do this is with $l=1$ currents on the shortest contour between $x$ and $y$. Therefore, the heavy currents induce an exponential decay  in the IR.

\section{Conclusion}
\label{sec:summary}

We have pointed to nonlinear sigma models in two dimensions at nonzero chemical potential because of two physically different transitions: the charge generation at $\mu_c$, the end-point of the Silver-Blaze region, where the symmetry gets broken down to the $O(2)$ subgroup (selected by $\mu$), and the Kosterlitz-Thouless (quasi)order-disorder transition at $\mu_{\text{KT}}$ driven by vortices/merons. Our motivation is to study effects that also occur in higher-di\-men\-sio\-nal gauge theories (asymptotic freedom, topological objects, nonzero density with sign problem etc.), but we believe that the interplay of these transitions is interesting in its own right. Individually, these transitions are of second and infinite order, respectively, but to our knowledge have not been combined in the same physical system. With ad hoc anisotropies one expects the KT transition, but not the Silver-Blaze phenomenon and the anisotropy of correlators, that we have discussed in Sec.~\ref{sec:O2}. On the other hand, for higher-
di\-men\-sio\-nal scalar theories at nonzero density, the KT transition is missing. We conjecture that $\mu_{KT}>\mu_c$, since the KT transition can only occur if the symmetry breaking forces the fields to be close to those of $O(2)$ that support vortices.
We have performed a large $N$-calculation to illustrate this point. Even if the two transition would coincide, one of them has to change its order. 

Using dualization techniques, we have also given lattice actions to study this system nonperturbatively, in the range where our methods are not reliable anymore. 

The two expected transitions could be further disentangled by hand through the use of terms proportional to ${\bm n}_\perp^2$ on top of the action at nonzero $\mu$ (breaking the fine-tuning between the terms linear and quadratic in $\mu$). From the large $N$ point of view, the introduction of order $N$ many chemical potentials could lead to novel effects.

Furthermore, it would be interesting to introduce complex $\mu$s into $O(3)$ and $CP(N)$ models, to see how this picture fits to the instanton constituents present at twisted boundary conditions (imaginary $\mu$). The latter can also be applied in spatial compactifications, after which the decompactification limit could be studied (in analogy to the $\mu=0$ case \cite{Dunne:2012ae}). $O(N)$ models do not contain topological objets to start with, but our mechanism using chemical potentials introduces them (albeit in the effective theory only), and they are presumably related to the mass gap generation.

Finally, let us comment on the relation between the setup we considered and QCD at 
nonzero baryon density (and zero temperature). At some $\mu_c$ at the order of the nucleon mass the Silver Blaze phenomenon ends and a (`liquid-gas') phase transiton to nuclear matter occurs. Whereas this effect is robust for all systems, other conjectured transitions are sensitive to the values of the quark masses, the electromagnetic interaction between them etc. Almost all model calculations, however, predict such transitions at $\mu>\mu_c$, just like we conjectured for the sigma models. 

The regime of asymptotically large chemical potentials is reliably computable and has been explored over a decade ago\footnote{although the idea of quark pairing at large quark densities is much older, see references in \cite{Alford:2007xm} for a review.}. It is color-superconducting and presumably in the color-flavor-locked phase \cite{Alford:1998mk} (depending on the number of light quark flavors which participate in the pairing mechanism, see \cite{Alford:2007xm} for a review), developing a diquark condensate which gives gauge fields a mass and acts much like a Higgs field. In this phase instantons do not suffer from the size moduli problem and are the dominant cause of the free energy`s $\theta$ dependence\footnote{provided, of course, the mass of the fermions is not identically zero, in which case there is no $\theta$ dependence.}. On the other hand, there were 
proposals in recent years that the phase of zero-temperature YM and QCD theories is related to the semiclassical regime of instanton-monopoles in related theories \cite{Unsal:2012zj,Poppitz:2012nz,Poppitz:2012sw,Poppitz:2013zqa,Anber:2014lba}, which is reflected in the $\theta$ dependence.  Independently, using large $N$ arguments, Zhitnitsky et al.\ suggested that instantons at the confinement/deconfinement transition  dissociate into ``instanton quarks'' in the confined phase \cite{Parnachev:2008fy,Zhitnitsky:2008ha}, while its similarity to the KT transition was discussed in \cite{Zhitnitsky:2013wfa}. This picture of instanton dissociation to some objects with fractional topological charge agrees with the recent lattice studies \cite{Bonati:2013tt,Bonati:2013dza}. As suggested by Zhitnitsky et al.,  QCD as a function of the chemical potential should undergo a similar transition from an instanton-dominated color superconductor phase at large $\mu$ to a phase of ionized instantons 
at smaller $\mu$. However, the transition is not expected to be a KT transition, which is characteristic to two dimensions. Nevertheless, the role of topological objects is substantially different in the two regimes and will be followed by a corresponding  change in the $\theta$-angle dependence, in qualitative agreement with the sigma model behavior discussed in this work.

\section{Acknowledgements}

The authors would like to thank Mohamed Anber, Christof Gattringer,  Erich Poppitz and Mithat \"Unsal for helpful discussions. This work has been supported by the DFG (BR 2872/6-1) and a scholarship of BayEFG.

\bibliographystyle{elsarticle-num}

\bibliography{bibliography}

\end{document}